\documentclass{aa}
\usepackage{graphicx}

\def\spose#1{\hbox to 0pt{#1\hss}}
\def\lta{\mathrel{\spose{\lower 3pt\hbox{$\mathchar"218$}}
        \raise 2.0pt\hbox{$\mathchar"13C$}}}
\def\gta{\mathrel{\spose{\lower 3pt\hbox{$\mathchar"218$}}
        \raise 2.0pt\hbox{$\mathchar"13E$}}}

\begin{document}

\thesaurus{06(02.01.2; 02.09.1; 08.02.1; 08.14.2)}

\title{The nature of dwarf nova outbursts}
\titlerunning{The nature of dwarf nova outbursts}
\authorrunning{Buat-M\'enard et al.}
\author{Valentin Buat-M\'enard\inst{1},
        Jean-Marie Hameury\inst{1}
        \and Jean-Pierre Lasota\inst{2}
}
\offprints{V. Buat-M\'enard}
\mail{buat@astro.u-strasbg.fr}
\institute{UMR 7550 du CNRS, Observatoire de Strasbourg,
           11 rue de l'Universit\'e, F-67000 Strasbourg, France,
           hameury@astro.u-strasbg.fr
           \and Institut d'Astrophysique de Paris,
           98bis Boulevard Arago, 75014 Paris, France,
           lasota@iap.fr
           }

\date{Received / Accepted}

\maketitle

\begin{abstract}
We show that if the dwarf-nova disc instability model includes the
effects of heating by stream impact and tidal torque dissipation in the
outer disc, the calculated properties of dwarf-nova outbursts change
considerably, and several notorious deficiencies of this model are
repaired. In particular: (1) outside-in outbursts occur for mass
transfer rates lower than in the standard model as required by
observations; (2) the presence of long (wide) and short (narrow)
outbursts with similar peak luminosities is a natural property of the
model. Mass-transfer fluctuations by factors $\sim$ 2 can explain the
occurrence of both long and short outbursts above the cataclysmic
variable period gap, whereas below 2 hr only short normal outbursts are
expected (in addition to superoutbursts which are not dealt with in this
article). With additional heating by the stream and tidal torques, such
fluctuations can also explain the occurrence of both outside-in and
inside-out outbursts in \object{SS Cyg} and similar systems. The
occurrence of outside-in outbursts in short orbital-period, low
mass-transfer-rate systems requires the disc to be much smaller than the
tidal-truncation radius. In this case the recurrence time of both
inside-out and outside-in outbursts have a similar dependence on the
mass-transfer rate $\dot{M}_{2}$.
\keywords
accretion, accretion discs -- instabilities -- (Stars:) novae, cataclysmic
variables -- (stars:) binaries : close
\end{abstract}

\section{Introduction}

Dwarf novae (DN) are erupting cataclysmic variable stars. The 4 - 6 mag
outbursts last from few days to more than a month and the recurrence
times can be as short as a few days and as long as 30 years (see e.g.
\cite{w95}). In cataclysmic variables, a Roche-lobe filling low-mass
secondary star is losing mass which is accreted by a white dwarf
primary. In DNs the matter transfered from the secondary forms an
accretion disc around the white dwarf; it is this disc which is the site
of the outbursts. It is believed that an accretion disc instability
due to partial hydrogen ionization is triggering the outbursts. The disc
instability model (DIM; see Cannizzo 1993, Lasota 2000 for reviews) is
supposed to describe the whole DN outburst cycle. This model assumes the
$\alpha$-prescription (\cite{shak73}) for the viscous heating and the
angular momentum transport in the disc. In such a framework, the
outbursts are due to propagating heating and cooling fronts, while during
quiescence the disc is refilling the mass lost during the outburst. In
the standard version of the model, the mass transfer rate is assumed to
be constant during the whole outburst cycle. In order to reproduce
outburst amplitudes and durations, the $\alpha$-parameter has to be larger
in outburst than in quiescence (Smak 1984).

The DIM is widely considered to be the model of DN outbursts because it
identifies the physical mechanism giving rise to outbursts, and with the
above mentioned assumptions it produces outburst cycles which roughly
look like the real thing. A closer look, however, shows that the model
fails to reproduce some of the fundamental properties of dwarf-nova
outbursts. The main deficiencies of the DIM were recently summarized by
Smak (2000). First, models predict a significant increase of the disc
luminosity during quiescence (see e.g. Fig. \ref{fig3}), an effect which
is not observed. Second, in many systems the distribution of burst
widths is bimodal (van Paradijs 1983). The DIM can reproduce the
width bimodality but, contrary to observations, its narrow outbursts
have lower amplitudes than the wide ones. Third, observations clearly
show that during outbursts the secondary is irradiated and the mass
transfer rate increases. The standard DIM does not take this into
account. To these problems pointed out by Smak one can add several other
deficiencies. The model predicts outbursts beginning in the outer disc
regions (`outside-in') only for high mass transfer while observations
show the presence of such outbursts at apparently low mass transfer
rates. In most cases the model predicts quiescent accretion rate which
are at least two orders of magnitude lower than the rates deduced from
quiescent luminosities. Finally, it is clear that superoutbursts
observed in SU UMa stars and other systems (Smak 2000) cannot be
explained by the standard DIM.

All this leads to the obvious conclusion that the standard version of
the DIM has to be enriched by inclusion of physical processes which
might be important but which, for various reasons, were not taken into
account in the original model. Osaki (1989) proposed that superoutbursts
are due to a tidal-thermal instability. Smak (2000) suggested that
problems of this model could be solved by a hybrid which would combine
it with the radiation-induced enhanced mass transfer model (Osaki 1985;
Hameury, Lasota \& Hur\'e 1997). He included mass transfer enhancements
during outburst into the standard model and suggested that this could
explain properties of the narrow and wide outbursts which alternate in
many systems (Smak 1999a). Hameury, Lasota \& Warner (2000), extending
previous work by Hameury et al. (1997) and Hameury, Lasota \& Dubus
(1999), showed that inclusion of effects such as mass transfer
variations, formation of holes in the inner disc and irradiation of both
the disc and the secondary allows to reproduce many properties of the
observed outbursts.

Some questions, however, were left without satisfactory answers. First,
the problem of `outside-in' outbursts occurring for too high
mass-transfer rates was still unsolved. Second, Smak (1999a) obtained
narrow and wide outbursts by including irradiation induced mass transfer
fluctuations but wide outbursts in his model result from fluctuations
whose amplitude bring the disc to a stable state. In such a case the
width of wide outbursts would not be determined by the disc structure
alone. There is nothing fundamentally wrong with such a model and long
standstills of Z Cam systems are most probably due to such fluctuations,
as proposed by Meyer \& Meyer-Hofmeister (1983) and shown by
Buat-M\'enard et al. (2000). However, one should check that the model is
complete, that no other mechanism can explain outburst width distribution.

In the present paper we study two mechanisms which dissipate energy in
the outer parts of an accretion disc. They are: (1) the impact on the
outer boundary of the accretion disc of the gas stream which leaves the
secondary through the $L_1$ critical point of the Roche lobe; this
impact zone is observed in the light-curves as a bright (`hot') spot;
(2) the viscous dissipation induced by the tidal torques which are
responsible for the disc outer truncation. These mechanisms which may
heat up the disc's outer regions have been known for a long time
(\cite{pacz77,pap-pring77,meyer83}), but have been neglected in most of
DIM calculations. The exception is Ichikawa \& Osaki (1992) who included
both effects in their model. They used, however, a viscosity prescription
specially designed to suppress inside-out outbursts and did not perform a
systematic study of the effects as a function of the mass-transfer rate,
size of the disc etc.

We add stream-impact and tidal torque heating effects to the Hameury et
al. (1998) version of the DIM, presented in Sect. 2. We use a 1D
approximation because the dynamical time-scale is shorter than the
thermal time-scale. We study the effect of the stream impact (Sect. 3)
and the tidal torques (Sect. 4) on outburst properties in short and long
period DN systems. Both effects change the energy equation of the
standard DIM. In both cases we find similar results: when compared to
the standard case, `outside-in' outbursts occur for lower mass transfer
rate. The recurrence time is decreased by both the stream impact and
dissipation of the tidal torques. We conclude, therefore,
that heating of the outer disc by the stream impact and the rate of work
by tidal torques provides a satisfactory explanation of the presence of
`outside-in' outbursts at realistic mass-transfer rates.

The most important result of our investigation is that when these
effects are included one can obtain two types of outbursts: short
(`narrow') and long (`wide') (Sect. 5). We find a critical mass transfer
rate for which these two type of outburst `alternate'. This mass
transfer rate is very close to that deduced from evolutionary
calculations for systems with orbital periods longer than 3 hours
(`above the period gap'). For lower mass-transfer rates only narrow
outbursts are present; for higher ones, outbursts are only of the wide
type. Fluctuations of the mass-transfer rate allow to expand the range
of orbital periods for which the bimodal outburst width distribution is
occurring. We conclude that the observed bimodal outburst width
distribution is caused by the outer-disc heating.

In Sect. 6 we discuss conditions for which outside-in outbursts can
occur. We show that one can naturally account for the occurrence of both
types of outbursts in \object{SS Cyg} if the mass transfer rate
fluctuates by factors $\lta 2$ around a mean value which is close to
what is generally assumed as the average mass transfer rate in this
system. Outside-in outbursts in systems below the period gap, which have
low mass transfer rates, can be explained only if the disc is
significantly smaller than the tidal truncation radius. We show that
this is in agreement with observations and consistent with expectations
of the tidal-thermal instability model, at least during the first half
of the supercycle of SU UMa systems. We also point out the ambiguity in
observational determination of outburst `outside-in' or `inside-out'
type, also respectively called A and B types. These types are generally
assigned to outbursts according to their shape and the UV delay. We
question the general validity of such criteria and argue that the
eclipse profile observations during outbursts is the only reliable
diagnostics of the outburst type.

\section{The model}

We use the recent version of the disc instability model described in
Hameury et al. (1998). We assume here that $\alpha=\alpha_{\rm
cold}=0.04$ in quiescence and $\alpha = \alpha_{\rm hot} = 0.2$ in
outburst (\cite{smak99b}). Difficulties in reproducing the observed UV
delays (see Sect. 6 for more details) and accounting for observed
quiescent X-ray fluxes well above the values predicted by the DIM, led
to the suggestion that `holes' are present in the center of quiescent
accretion discs (\cite{meyer94,lasota95}). Smak (1998) pointed out,
however, that the failure to reproduce observed UV delays by numerical
codes was largely due to the assumption of a fixed outer disc radius,
which prevented the occurrence of outside-in outbursts. In addition, the
discs used in such calculations were too small to give the required
time-scales. When one uses reasonable outer boundary conditions, long UV
delays can be observed. However, if one wishes to reproduce both the
delays {\em and} the light-curve shapes, a truncated disc might be still
a necessary ingredient of the model (Hameury et al. 1999). Such a hole
could be due to magnetic disruption in the case of a magnetic white
dwarf or to evaporation in the general case. Because we are not
interested here in the exact values of the rise time and the UV delay,
we use a fixed size hole only for numerical convenience. The inner disc
radius is fixed at $r_{\rm in}=10^9$ cm, while $5 \times 10^8$ cm is
approximately the radius of a 1.2 M$_\odot$ white dwarf and $6.9 \times
10^8$ cm correspond to the other mass we use: 0.8 M$_\odot$. The outer
boundary condition depends on the circularization radius $r_{\rm circ}$
at which a particle that leaves the Lagrangian point $L_1$ with angular
momentum $j = \Omega_{\rm orb} b^2$ (where $\Omega_{\rm orb}$ is the
orbital angular velocity and $b$ the distance between the primary and
the $L_1$ point), would stay in circular orbit if there were no
accretion disc. From $j = \Omega_{\rm K} r_{\rm circ}^2$, where
$\Omega_{\rm K}$ is the Keplerian angular frequency, one gets: $r_{\rm
circ} \simeq \Omega_{\rm orb}^2 b^4 / G M$. When one takes the
gravitational potential of the secondary into account, these relations
are slightly modified, and we linearly interpolate the table from Lubow
\& Shu (1975) to calculate $r_{\rm circ}$. The mean outer radius $<
r_{\rm out} >$ is taken to be the average of the three radii $r_1$,
$r_2$ and $r_{\rm max}$ calculated in table 1 of Paczy\'nski (1977).

We consider in this paper the effects of the impact of the gas stream
coming from the secondary (Sect. 3) and of the dissipation induced by
tidal torques (Sect. 4) in both short and long period dwarf novae (thus
for both small and large accretion discs) using two characteristic sets
of parameters taken from Ritter \& Kolb (1998): those of \object{SU UMa}
and \object{SS Cyg} (Table \ref{table1}). For the short period systems,
although we take the parameters of SU UMa, we include neither the tidal
instabilities (suggested by Osaki 1989) to be responsible for
superoutbursts in SU UMa), nor the combination of irradiation, mass
transfer variation and evaporation used by Hameury, Lasota \& Warner
(2000) to obtain long lasting outbursts. Therefore, we do not expect to
reproduce superoutbursts. We also neglect irradiation of secondary star
which in principle may have significant impact on some of our results.
We will consider this effect in a future work.

\begin{table}
\caption{\label{table1} Parameters adopted for long and short period
systems.}
\begin{tabular}{lcc}\hline
 & Long period & Short Period\\
\hline
$M_1 / $M$_\odot$ & 1.2 & 0.8\\
$M_2 / $M$_\odot$ & 0.7 & 0.15\\
$P_{\rm orb}$ (hr) & 6.6 & 1.83\\
$r_{\rm circ}$ ($10^{10}$ cm) & 1.14 & 0.91\\
$< r_{\rm out} >$ ($10^{10}$ cm) & 5.4 & 2.3\\
\hline
\end{tabular}
\end{table}

\section{Stream impact}

\subsection{Description}

The gas stream from the secondary hits the accretion disc creating the
observed bright spot. As the dynamical time is smaller than the thermal
time ($t_{\rm dyn} / t_{\rm th} = \alpha < 1$), the temperature is
roughly constant around each annulus of the accretion disc. Thus, the effect
of stream impact could be approximatively axisymmetric and we can use a
1D approximation. This assumption is obviously very rough. Spruit \&
Rutten (1998) have shown that the stream impact region in \object{WZ
Sge} is not at all axisymmetric (see also Marsh \& Horne 1990 for
\object{IP Peg}). Moreover, if we refer to the generally assumed values
of $\alpha$, $t_{\rm dyn} / t_{\rm th}$ is not that small in outburst;
in addition, as the hot spot temperature $T_{\rm hs} \gg T_{\rm disc}$
in the outer layers, the actual thermal time is much smaller than the
unperturbed equilibrium value. Finally, we do not know exactly how the
stream impact interacts with the disc and where precisely its energy
deposited in the disc; e.g. in the simulations of Armitage \& Livio
(1996, 1998) a significant amount of the stream material can ricochet
off the disc edge and overflow toward smaller radii.

It is therefore difficult to model properly the hot-spot contribution to
the energy balance of the outer layers of an accretion disc. Here we
consider that the stream impact heats an annulus $\Delta r_{\rm hs}$ of
the disc with an `efficiency' $\eta_{\rm i}$. The coefficient $\eta_{\rm
i}$ is supposed to represent the rather complex way the outer disc is
heated by the stream.
The energy per unit mass released in the shock between the stream and the disc is
$({\bf V_2 - V_1})^2$, where $\bf V_2$ is the keplerian velocity at the outer
disc edge, $|{\bf V_2}| = G M_1/(2r_{\rm out})$, and $\bf V_1$ is the stream
velocity at the position of impact. As the angle between $\bf V_1$ and $\bf
V_2$ is large, and $|{\bf V_1}|$ is comparable to $|{\bf V_2}|$, $|{\bf V_2 -
V_1}|$ is of order of $G M_1/(2r_{\rm out})$.

The heating rate $Q_{\rm i}$ is therefore taken as:
\begin{equation}
Q_{\rm i}(r) = \eta_{\rm i} \frac{G M_1 \dot{M_2}}{2 r_{\rm
out}} \frac{1}{2 \pi r_{\rm out} \Delta r_{\rm hs}}
\exp \left(-\frac{r_{\rm out}-r}{\Delta r_{\rm hs}}\right)
\label{eq:qi}
\end{equation}
where $M_1$ is the primary's mass, $\dot{M_2}$ the mass transfer rate from the
secondary and $r_{\rm out}$ the accretion disc outer radius. This assumes
that the difference between the stream and the Keplerian kinetic energy is
released in a layer of width $\Delta r_{\rm hs}$ with an exponential
attenuation. Note that $\eta_{\rm i}$ can be larger than unity, as $|{\bf V_2
- V_1}|$ can be larger than the Keplerian velocity. In the following, we take
$\eta_{\rm i}$ = 1.

The energy equation then becomes:
\begin{eqnarray}
\frac{\partial T_{\rm c}}{\partial t} = \frac{2(Q^+ + 1/2 Q_{\rm i} - Q^- +
J)} {C_P \Sigma} - \frac{\mathcal{R} T_{\rm c}}{\mu C_P} \frac{1}{r}
\frac{\partial r v_r}{\partial r} \nonumber \\
- v_r \frac{\partial T_{\rm c}}{\partial r}
\end{eqnarray}
Here, $T_{\rm c}$ is the central temperature, $Q^+ = (9/8) \nu \Sigma
\Omega_{\rm K}^2$ and $Q^- = \sigma T_{\rm eff}^4$ the heating and
cooling rate respectively, $\Sigma$ the surface density, $\nu$ the
kinematic viscosity coefficient, $J$ is a term that accounts for the
radial energy flux carried either by viscous processes or by radiation
(see Hameury et al. 1998), and $v_r$ is the radial velocity. $C_P$ is the
specific heat at constant pressure, $\mu$ the mean molecular weight, and
$\mathcal{R}$ the perfect gas constant.

The cooling rate $Q^-(\Sigma , T_{\rm c}, r)$ is obtained from
interpolation in a grid obtained by solving the vertical structure of
the disc (\cite{jmh98}). These structures are calculated assuming a
steady state, the departure from thermal equilibrium being accounted for
by an effective viscosity parameter $\alpha_{\rm eff}$ different from
the actual viscosity parameter $\alpha$. We assume here that the effect
of the stream impact on the vertical structure can also be accounted for
in this way. This would be perfectly correct if the local heating term
were proportional to pressure, as is viscosity; if on the other hand it
were localized at the disc surface, the problem would resemble that of
disc illumination, and also in such a case, the use of an effective
$\alpha_{\rm eff}$ leads to qualitatively correct results: Stehle \&
King (1999), who used this approximation, and Dubus et al. (1999), who
calculated the exact vertical structure of irradiated discs, obtained
similar modifications of the S curves by irradiation of the disc.

\subsection{Results}

\begin{figure}
\resizebox{\hsize}{!}{\includegraphics{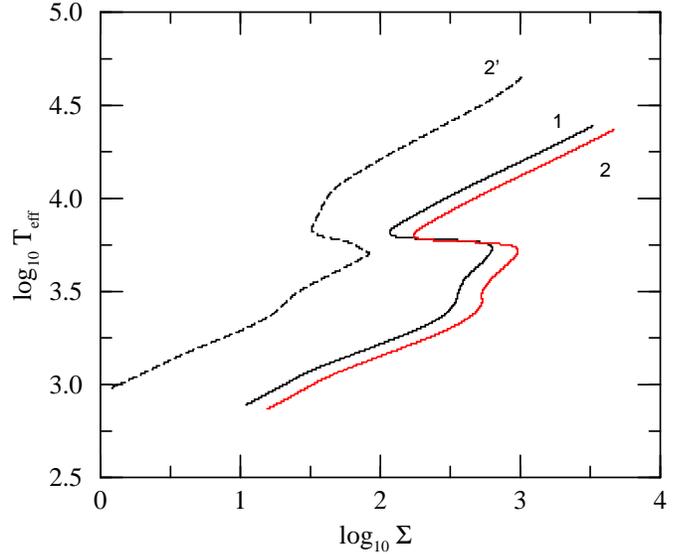}}
\caption{$\Sigma - T_{\rm eff}$ curves for the parameters of SS Cyg . Curve 1 is
calculated at a radius $ 3.75 \times 10^{10}$ cm that is out of the stream
impact region. Curves 2$^{'}$ and 2 are calculated at a radius 5.2
10$^{10}$ cm with and without the stream impact included in the model.
For the hot branch $\alpha_{\rm hot}=0.2$,  for the cold one $\alpha_{\rm cold}
=0.04$. The S-curve is obtained using Eq. (40) in Hameury et al. 1998}
\label{fig1}
\end{figure}

Fig. \ref{fig1} displays the $\Sigma - T_{\rm eff}$ curves, calculated at two
different radii with and without the heating by the stream impact. These are
the well known 'S-curves' in which three branches can be distinguished: the
stable, low- and high-state branches where $T_{\rm eff}$ increases with
$\Sigma$, and the unstable middle branch where $T_{\rm eff}$ decreases with
$\Sigma$ increasing. The latter branch is delimited by two critical points $(\Sigma_{\rm
min},T_{\rm min})$ and $(\Sigma_{\rm max},T_{\rm max})$. To compute these
S-curves, a steady state was assumed so $T_{\rm eff}$ is proportional to
$\dot{M}^{1/4}$. As mentioned above, we take $\eta_{\rm i}$ = 1.

\begin{figure}
\resizebox{\hsize}{!}{\includegraphics{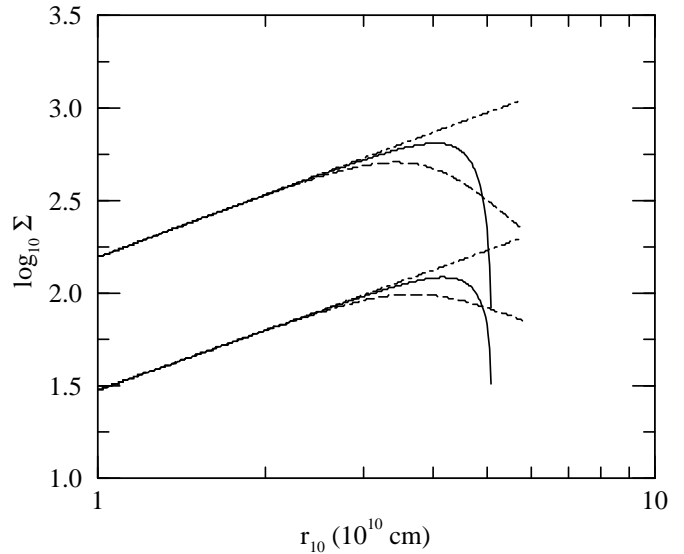}}
\caption{$\Sigma_{\rm min}$ (lower curves) and $\Sigma_{\rm max}$ (upper
curves) for SS Cyg parameters. The dot-dashed lines show the results of
the standard DIM, the solid lines have been obtained when the stream
impact is taken into account, and the dashed lines when tidal effects
are included.} \label{fig2} \end{figure}

When stream impact is taken into account one observes at large radii
significant deviations from the standard case. It must, however, be
stressed that $Q_{\rm i}$ is a non-local quantity proportional to
$\dot{M}_2$ and thus depending on the outer boundary condition, whereas
the usual S-curves are locally defined. Hence, the interpretations of
these curves is not straightforward. For a fixed $\Sigma$, there is a
factor $\Delta \log T_{\rm eff} \sim 0.5$ between curves 2 and 2$^{'}$.
For a fixed $T_{\rm eff}$, $\Sigma$ is reduced in the outer disc region
when heating by the stream impact is included. It should therefore be
easier, for a fixed mass transfer rate, to trigger outbursts starting at
the outer edge of the disc when heating by the stream impact is
included. One also expects the disc to be less massive. Moreover, the
$\Sigma_{\rm min} - \Sigma_{\rm max}$ interval decreases (see Fig.
\ref{fig2}) so that the recurrence time should decrease too.

\begin{figure}
\resizebox{\hsize}{!}{\includegraphics[angle=-90]{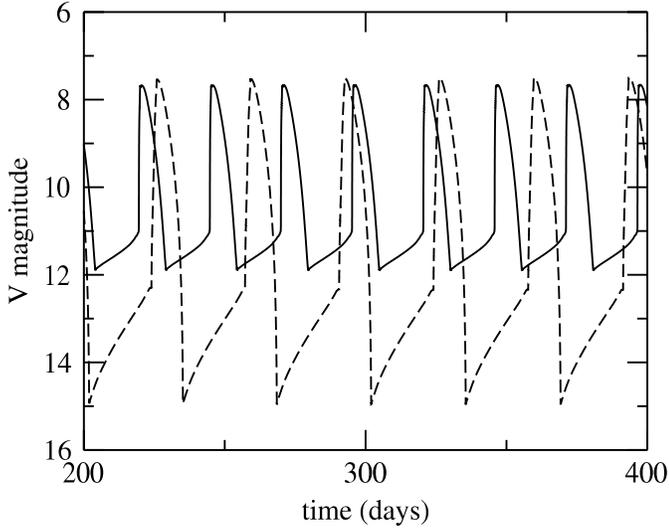}}
\caption{Comparison of light curves with (solid line) and
without the stream impact effect (dashed line). The parameters we use are
taken from table \ref{table1} and $\dot{M}_2 = 2 \times 10^{17}$ g s$^{-1}$,
$\eta_{\rm i} = 1$ and $\Delta r_{\rm hs} = r_{\rm out} / 10$.}
\label{fig3}
\end{figure}

\begin{figure}
\resizebox{\hsize}{!}{\includegraphics{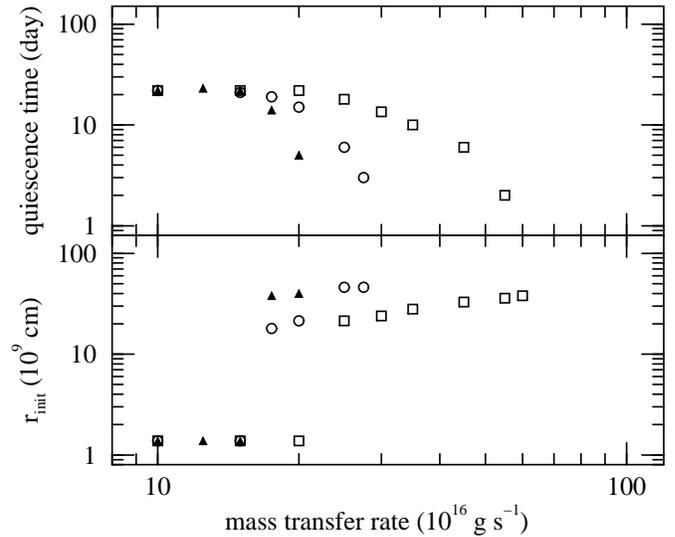}}
\caption{Radius $r_{\rm init}$ at which the instability is triggered and
quiescence time as a function of the mass transfer rate from the secondary.
The parameters are those of SS Cyg. Symbols correspond to the following:
squares for the standard DIM; circles when the stream impact is included;
triangles when the stream impact and the tidal dissipation are both included.
The high mass-transfer end of each sequence of symbols corresponds approximately
to the stability limit.}
\label{fig4}
\end{figure}

\begin{figure}
\resizebox{\hsize}{!}{\includegraphics{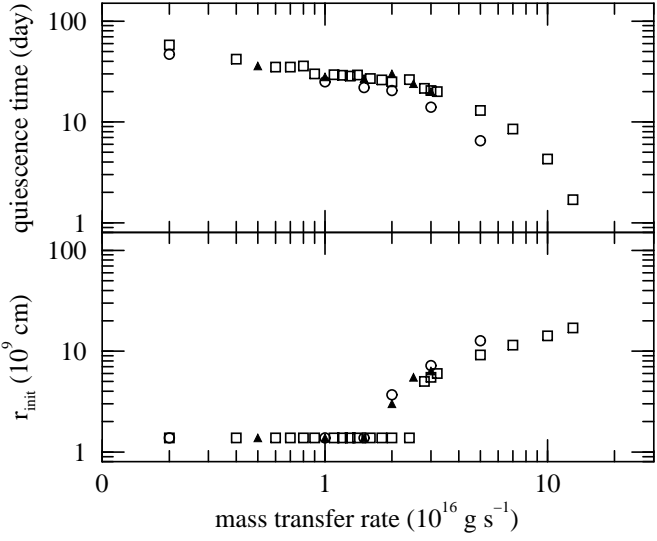}}
\caption{Same as Fig. \ref{fig4} for short period systems}
\label{fig5}
\end{figure}

Fig. \ref{fig3} shows light curves calculated without and with heating
for a long period system (``SS Cyg"). One can see that the recurrence
time is reduced in the model with the stream impact and that the
quiescent luminosity is increased, as expected, since the stream impact
heats up the outer parts of the disc which dominate the quiescent
luminosity. During quiescence, however, the luminosity still increases
in contrary to observations. Figures \ref{fig4} and \ref{fig5} show how
the outburst type and the recurrence times depend on the mass-transfer
rate. Including the additional heating by the stream impact implies that
outside-in outbursts are triggered for lower mass transfer rates than in
the standard DIM. Also the upper critical rate $\dot{M}_{\rm c}$, above
which the disc is in a steady hot state, is lower as predicted by Meyer
\& Meyer-Hofmeister (1983).

For larger $\Delta r_{\rm hs}$, the quiescent luminosity is smaller,
because the additional amount of energy per unit surface decreases in
the outer parts of the disc (note also that the integral of $Q_{\rm i}$
over the disc surface depends weakly on $\Delta r_{\rm hs}$ because of
the simplified form of Eq. \ref{eq:qi} where terms of order of $\Delta
r_{\rm hs} / r_{\rm out}$ have been neglected). For smaller $\eta_{\rm
i}$, the recurrence time and the quiescent magnitude are higher. As
short period systems have small discs, one might have expected that the
stream impact would have stronger effects in these systems, but the
results are really similar as can be seen in Figs. \ref{fig4} and
\ref{fig5}. However, in these calculations we assumed that the average
disc outer radius is equal to the tidal radius. This is an overestimate
of the disc's size, especially for the SU UMa stars. Realistically small
discs will be discussed in Sect. \ref{oi}.

The main effects of the stream-impact heating seen on these figures are:
(1) The critical accretion rate $\dot{M}_{\rm c}\left(\Sigma_{\rm
min}\right)$ above which the disc is stable, is decreased by a factor of
more than 3 ; (2) the transition between inside-out and outside-in
outburst ('A-B transition') occurs for a transfer rate $\dot{M}_{\rm
A-B}$ that is decreased by a factor $\leq 2$ ; (3) the recurrence time
is almost unchanged. As expected (Osaki 1996), the quiescence time is
approximately constant when the outbursts are of the inside-out type,
and decreases (very roughly as $\dot{M}_2^{-2}$) for outside-in
outbursts, although for the smaller disc some deviations from this
relation can be seen (see Sect. 6). (Note that we plot here the time in
quiescence and not the recurrence time.)

For systems with parameters close to that of SS Cyg, the inclusion of
the stream impact heating lowers the critical rate for outside-in
outburst from $\dot{M}_{\rm A-B} \simeq 2.5 \times 10^{17}$ g s$^{-1}$
to $\dot{M}_{\rm A-B} \simeq 1.7 \times 10^{17}$ g s$^{-1}$ . The
quiescence time is also shorter but only by a few days as long as
$\dot{M}_2$ is not too close to $\dot{M}_{\rm c}$ (see Fig. \ref{fig4}).

For short period systems, the critical mass transfer is $\dot{M}_{\rm c}
\simeq 5 \times 10^{16}$ g s$^{-1}$, however, the transition between
inside-out and outside-in outbursts still occurs for mass-transfer rates
($ \dot{M}_{\rm A-B} \simeq 2 \times 10^{16}$ g s$^{-1}$) much higher
that what is estimated for these systems (less than a few $ \times
10^{15}$ g s$^{-1}$; see e.g. Warner 1995). We will discuss this problem
in Sect. \ref{oi}.

\section{Tidal dissipation}

\subsection{Description}

Accretion discs in close binary systems are truncated by the tidal
effect of the secondary star at a radius close to the Roche radius, the
so-called tidal truncation radius: angular momentum transported outwards
by viscous torques is returned to the binary orbital motion by the tidal
torques acting on the disc's outer edge. In the standard DIM model,
tidal torques can be included in the angular momentum equation in the
following way (Smak 1984; Hameury et al. 1998):
\begin{equation}
\frac{\partial \Sigma}{\partial t} = - \frac{1}{r} \frac{\partial}{\partial r} (-
\frac{3}{2} r^2 \Sigma \nu \Omega_{\rm K}) - \frac{1}{2 \pi r} T_{\rm tid}
\end{equation}
where the tidal torque per unit of area can be written as
\begin{equation}
T_{\rm tid} = c r \nu \Sigma \Omega_{\rm orb} \left(\frac{r}{a}\right)^n
\label{eq:tid}
\end{equation}
where $a$ is the orbital separation of the binary and $c$ a numerical
constant that allows to adjust the tidal truncation radius $< r_{\rm
out} >$ that the disc would have had if it were steady (this gives an
average value of the outer radius). The $r^n$ term used in this formula
determines the disc fraction that is significantly affected by the
tidal-torque dissipation (it corresponds to radii where the $\Sigma_{\rm
min}$ and $\Sigma_{\rm max}$ curves are decreasing with radius (Fig.
\ref{fig2})). We use $n=5$ which is quite an uncertain value, and we
discuss in Sect. 6 the effects of changing it.

As pointed out by Smak (1984) and discussed in detail by Ichikawa \& Osaki
(1994), tidal torques (as viscous torques) induce a viscous dissipation
(\cite{pap-pring77}):
\begin{equation}
Q_{\rm tid} = (\Omega_{\rm K} - \Omega_{\rm orb}) T_{\rm tid}
\end{equation}
This tidal dissipation is usually neglected as it is small compared to $Q^+$
except near the outer radius. During outbursts, the disc radius varies
significantly, so does the tidal-torque dissipation which can be large enough
to modify the outburst behaviour. We therefore include this term in the energy
equation:
\begin{eqnarray}
\frac{\partial T_{\rm c}}{\partial t} = \frac{2(Q^+ + 1/2 Q_{\rm tid} + 1/2
Q_{\rm i} - Q^- + J)}{C_{\rm P} \Sigma} \nonumber & \\
- \frac{\mathcal{R} T_{\rm c}}{\mu C_{\rm P}}
\frac{1}{r} \frac{\partial r v_{\rm r}}{\partial r}
 - v_{\rm r} {\partial T_{\rm c} \over \partial r} &
\end{eqnarray}

\subsection{Standard DIM plus tidal dissipation}

We first discuss the tidal effect alone, without including the heating by the
stream impact. We find that type A outbursts are obtained for lower mass
transfer rates than in the standard case; $\dot{M}_{\rm A-B}$ is only
slightly larger than in the case where the stream impact is included. For
long period systems, we found $\dot{M}_{\rm A-B} \simeq 1.8 \times 10^{17}$g
s$^{-1}$. The critical mass transfer rate $\dot{M}_{\rm c}$ is also slightly
larger than in the model with stream impact.

\begin{figure}
\resizebox{\hsize}{!}{\includegraphics[angle=-90]{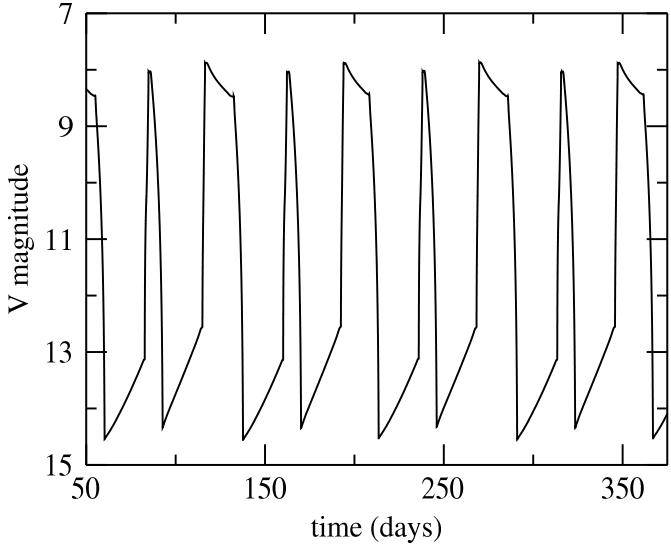}}
\caption{Light curve for the DIM plus tidal dissipation
(without stream impact) for $\dot{M}_2 = 1.5 \times 10^{17}$g s$^{-1}$
and binary parameters of SS Cyg. The
Both short (narrow) and long (wide) outbursts are of inside-out
type.}
\label{fig6}
\end{figure}

The major difference with previous calculations is the presence of a critical
mass transfer rate $\dot{M}_{\rm SL}$, for which one obtains a sequence of
alternating short (narrow) and long (wide) outbursts with {\em roughly the 
same amplitudes} (see Fig. \ref{fig6}). At low mass transfer rates only short
outbursts are present whereas for mass transfer rates higher than 
$\dot{M}_{\rm SL}$ all outbursts are long. In a small range of mass transfer 
rates outbursts of both durations are present (see Fig. \ref{fig7}). We
return to this point in the next section.

$\dot{M}_{\rm SL}$ is comparable to $\dot{M}_{\rm A-B}$; for the
parameters used in Fig. \ref{fig6}, $\dot{M}_{\rm SL} < \dot{M}_{\rm
A-B}$, and short outbursts are all of the inside-out type. They are
similar to those produced in the standard model, except that the
quiescent flux is slightly larger as a result of viscous dissipation in
the outer disc (as in the case of stream impact). As for the stream-impact 
heating case, the outburst peak luminosity is very slightly lower than in the 
standard model for the same mass transfer rate. The recurrence time scales 
are not affected by tidal dissipation when only short outbursts are produced. 
Long outbursts are slightly brighter than the short ones as observed 
(\cite{oppen98}), and for this set of parameters they can be either of the 
inside-out or outside-in type because the critical mass-transfer rate above 
which they occur $\dot{M}_{\rm SL}$ is less than $\dot{M}_{\rm A-B}$.

Long outbursts appear because near the outburst maximum the outer radius
increases, so does tidal dissipation. Tidal dissipation lowers the value
of $\Sigma_{\rm min}$ so that it takes longer to reach this critical
density at which a cooling-front starts to propagate and shuts off the
outburst. Outbursts therefore last longer than in the standard case and
have the characteristic ``flat top" shape. The amplitude of the
outer-disc variations increases with the mass-transfer rate. At low
mass-transfer rate these variations are too small to produce an
observable effect on the outburst properties. The critical rate
$\dot{M}_{\rm SL}$ corresponds to the case when the disc expansion
begins to affect the outburst duration. Short outburst follow long ones
because the tidal dissipation lowers both the critical and the actual
post-outburst surface-density so that during the next outburst the time
to empty the disc is shorter. Sequences of alternating short and long
outbursts are present for rather narrow interval of mass-transfer rates.
At higher mass-transfer rates outbursts are of the outside-in type and
therefore long (see Hameury et al. 1998).

Compared to the standard case, the recurrence time is increased when
long outbursts are present, but the duration of quiescence is unchanged:
the duration of the outburst is the only modified quantity.

When stream impact heating and the tidal dissipative process are both
included in the model, their effects cumulate: $\dot{M}_{\rm SL}$,
$\dot{M}_{\rm A-B}$ and $\dot{M}_{\rm c}$ are smaller by about 15 \% than
what one gets when including only one effect (see Fig. \ref{fig4} for
$\dot{M}_{\rm A-B}$ and $\dot{M}_{\rm c}$).

\section{The nature of long and short outbursts}

We have therefore a natural explanation of the coexistence of short
and long outbursts in long period ($>$ 3 hr) systems: for which the mass
transfer is large enough for the additional heating to produce this effect;
in short period system, $\dot{M_2}$
corresponds to short outbursts only. Heating by stream impact and
tidal-torque dissipation account also for the occurrence of outside-in
outbursts in \object{SS Cyg}, provided that the mass transfer rate
happens to be large enough (approximatively two to three times larger
than average).

In many cases, a bimodal distribution of the outbursts is observed
(\cite{vanP83}). In the case of \object{SS Cyg}, Cannizzo \& Mattei (1992)
identified short outbursts with a duration between 5 and 10 days, and long
ones with a duration between 13 and 17 days. Cannizzo (1993) obtained in his
model a sequence of alternating short and long outbursts, but this was due to
the use of a fixed outer radius which is an incorrect (`brick wall') boundary
condition (\cite{jmh98,smak98}). In any case his sequence looks different
from the observed one and all his outbursts are inside-out (type B).

When tidal dissipation is included, one obtains, for a SS Cyg-like system, an
alternating sequence of narrow and wide outbursts for a mass transfer rate
$\dot{M}_2 \sim \dot{M}_{\rm SL} \sim 1.5 \times 10^{17}$g s$^{-1}$ (the
value used in Fig. \ref{fig6}). As the short/long sequence of outbursts is
not uncommon, a small mass transfer rate variation around $\dot{M}_{\rm SL}$
is the most likely cause for the occurrence of both narrow and wide
outbursts. Moreover, if one believes that both A and B outburst types occur
in \object{SS Cyg} (or another long period system), $\dot{M}_2$ must also
vary around $\dot{M}_{\rm A-B}$.

\begin{figure}
\resizebox{\hsize}{!}{\includegraphics[angle=-90]{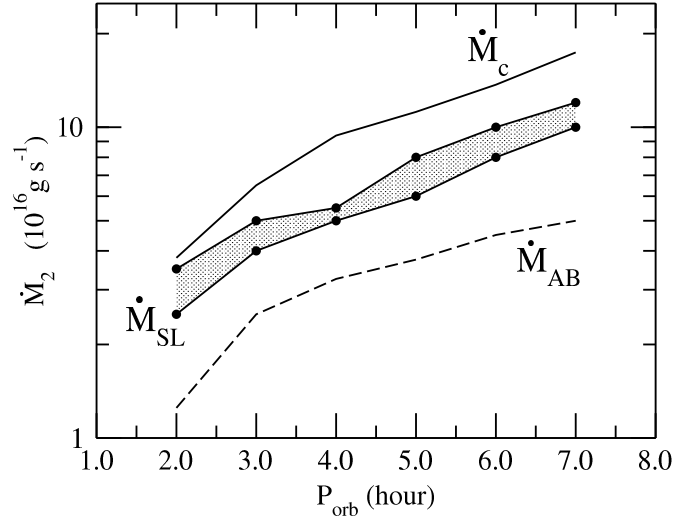}}
\caption{Range of mass transfer rates for which both long and short outbursts
occur and critical rate, as a function of the orbital period. Here, $M_1 =
0.6 \; $M$_{\odot}$ and $M_2 = 0.11 \times P_{\rm orb} \;$M$_{\odot}$. The
critical rate $\dot{M}_{\rm c} = 9.5 \times 10^{15} \dot{M}_1^{-0.89}
r_{10}^{2.68}$ (see \cite{jmh98}). For low $\dot{M}$, only short outbursts
are observed; above the upper curve, all outbursts are long. The upper solid
line shows the critical mass transfer rate above which the disc is stable.}
\label{fig7}
\end{figure}

The value of $\dot{M}_{\rm SL}$ depends on the binary parameters and for
a given orbital period one obtains a small range of mass transfer rates
for which both short and long outbursts occur. Fig. \ref{fig7} shows
this range as a function of the orbital period for a standard dwarf nova
with a primary mass $M_1 = 0.6 $M$_\odot$ and a main-sequence secondary
mass $M_2 = 0.11 \times P_{\rm hr} \;$M$_\odot$. Note that now the
primary's mass is different from the values used previously. $M_1 = 0.6
$M$_\odot$ corresponds to the average white-dwarf mass in CVs. As
mentioned earlier, $< r_{\rm out} >$ is taken from a table in
Paczy\'nski (1977). The lower boundary of this range is $\dot{M}_{\rm
SL}$ which varies by a factor of about 4 from a 2 hour period to a 7
hour period system. For short period systems, the mass transfer rate
required to obtain both narrow and wide outburst is larger than $2
\times 10^{16}$g s$^{-1}$. This is much larger than the expected secular
values of a few times $10^{15}$g s$^{-1}$ and can explain why narrow and
wide outburst alternation is not observed in short period dwarf novae.
Superoutbursts in SU UMa's, which are different from long outbursts
(Warner 1995; Smak 2000), occur for mass transfer rates lower than
$\dot{M}_{\rm SL}$. On the other hand there exist a subclass of the SU
UMa stars, the ER UMa systems, which have very short supercycles,
and are in outburst most of the time. They are presumed to have mass
transfer rates as large as $10^{16}$g s$^{-1}$, of order of or above the
value of $\dot{M}_{\rm SL}$ corresponding to their orbital periods, and
should therefore show the alternation of long and short outbursts
described here. This effect could add to the enhancement of the
mass-transfer rate resulting from the illumination of the secondary that
also cause sequences of long and short outbursts in short period systems
(Hameury et al. 2000).

Fig. \ref{fig7} shows also $\dot{M}_{\rm A-B}$; in this case we find
that $\dot{M}_{\rm SL}$ is larger than $\dot{M}_{\rm A-B}$, which means
that all long outbursts are now of the outside-in type. However, the resulting
light curves in which an alternance of short and long outbursts is found do not
differ much from the one shown in Fig. \ref{fig6}, except that the rise times
are shorter.

We propose that stochastic variations of the mass-transfer rate for the
secondary cause the sequence of alternating short and long, as well as
type A and type B outbursts. For the parameters of SS Cyg, we need 
mass-transfer rate variations in the range $\sim 1 - 2 \times 10^{17}$ g
s$^{-1}$ to obtain the four types in the same light curve; this range
depends on the system parameters, and in particular on the orbital
period (see Fig. \ref{fig7}), and on the mass of the primary. Because
outburst duration depends mainly on the disc's size (Smak 1999b) for
short outbursts whose duration is independent of the mass transfer rate
we obviously reproduce the correlation (van Paradijs 1983) between the
outburst width and the orbital period. It is much more difficult to
reproduce the observed correlation for long outbursts, since their
duration is quite sensitive to $\dot{M}_2$, whose variation with the
orbital period is not well known. One must, however, be cautious when
interpreting this correlation, as (i) the statistics is small, and (ii)
this correlation may result in part from the very definition of ``long"
outbursts: they must last longer than short outbursts, whose duration
increases with the orbital period simply because of the increase of the
disc's size (Smak 1999b). In the model by Smak (1999a), long outbursts
have to be interrupted by $\dot{M}_2$ returning to a low value since
otherwise the disc would remain steady. This requires the action of an
external physical mechanism with an unspecified free parameter. If for
long outbursts the correlation is real, in Smak's model the mechanism
terminating long outbursts would have to depend on the orbital period.

Since $\dot{M}_{\rm SL}$ is much larger than the mass-transfer rates
expected in systems below the period gap, we do not expect long
outbursts for periods shorter than 2 hours. This agrees with
observations. Superoutbursts observed in SU UMa systems must have a
different explanation; they are either due to a tidal instability, as
proposed by Osaki (1996), and/or to the irradiation of the secondary,
which is quite efficient in these short period systems (Hameury et al.
2000).

We still cannot, however, account for the presence of outside-in outbursts
in short period systems, whereas normal outbursts in SU UMa systems are
generally believed to be of this type (Osaki 1996). There are two possible
ways out of this difficulty: first, as discussed in the Sect. 6, SU UMa
normal outbursts occur in truncated discs and could be of the inside-out type;
second, the disc outer radius could be much smaller than what the standard
DIM predict (Smak 2000); indeed, observations of eclipsing dwarf novae by
Harrop-Allin and Warner (1996) tend to substantiate this idea. The
second option is also favoured by the `direct' observations of outside-in
outbursts in two eclipsing SU UMa systems (see Sect. \ref{obs} for
references).

\section{Inside-out versus outside-in outbursts}
\label{oi}

Outbursts of SS Cyg  are believed to be of both A and B types 
(\cite{smak84,can86, mauche96,smak98}). Thus, the mass transfer rate
from the secondary probably varies around the A-B transition. In models,
outside-in outbursts are usually obtained only for mass transfer rates
large compared to those expected from binary evolution models
(\cite{bk1999}). Moreover, disc model computations with such values of
$\dot{M}_2$ predict recurrence times that are much shorter than
observed. Ichikawa \& Osaki (1992) managed to get rid of this problem by
using an $r$-dependent $\alpha$-prescription specially designed to
suppress type B outbursts. Their viscous diffusion time is then almost
independent of the radius and inside-out outbursts occur only at very
small mass transfer rates.

As mentioned earlier, including the stream-impact heating 
lowers $\dot{M}_{\rm A-B}$ without changing the $\alpha$-prescription. We
obtain $\dot{M}_{\rm A-B} = 1.7 \times 10^{17}$ g s$^{-1}$, 1.4 times
smaller than in the standard case; this is reasonably close to estimates
of SS Cyg average mass transfer rate $\dot{M}_2 = 6 \times 10^{16}$g s$^{-1}$
(e.g. \cite{pat84}). An increase of the mass transfer rate by a factor 2
-- 3 is therefore sufficient to provoke outside-in outbursts. On the
other hand, for short period systems, the transition between inside-out
and outside-in outburst still occurs for much too large mass transfer
rates ($2 \times 10^{16}$ g s$^{-1}$), whereas estimates of mass transfer
in these systems do not exceed a few $ \times 10^{15}$ g s$^{-1}$
(Warner 1995; \cite{bk1999}).

\subsection{Predictions of the DIM}

Outside-in outbursts occur if the time scale for matter accumulation at the
outer disc edge $t_{\rm accum}$ is shorter than the viscous drift time scale
$t_{\rm drift}$. These times are (Osaki 1996):
\begin{equation}
t_{\rm accum} = {4 \pi^2 r^2 \nu \Sigma_{\rm max}^2(r) \over \dot{M}^2} \\
\label{tacc}
\end{equation}
\begin{equation}
t_{\rm drift} = r^2 \delta \nu^{-1}
\end{equation}
where $\nu = 2/3 \alpha (\mathcal{R }/\mu) T/\Omega$ is the kinematic
viscosity, and $\delta \sim 0.05$ is a numerical correction factor. Using the
analytical fits of $\Sigma_{\rm max}$ given by Hameury at al. (1998), the
condition for outside-in outbursts is:
\begin{equation}
\dot{M} > 4.35 \times 10^{15} \; \alpha_{\rm c}^{0.17} \left( {T \over 4000 \; \rm K}
\right) M_1^{-0.88} r_{10}^{2.64} \; \rm g \; s^{-1}
\label{eq:oi}
\end{equation}
where $r_{10}$ is the radius measured in 10$^{10}$ cm and where we have
assumed $\mathcal{R}/\mu = 5 \times 10^7$. This condition can be compared
with our numerical results. In the case without the additional effects, our
critical rate is $2.5 \times 10^{17}$ g s$^{-1}$ for long period systems,
whereas Eq. (\ref{eq:oi}) gives $1.8 \times 10^{17}$ g s$^{-1}$; for short
period systems, these two numbers are $2.5 \times 10^{16}$ and $2.8 \times
10^{16}$ g s$^{-1}$ respectively. It must be noted that the quality of the
agreement should not be a surprise, since the quantity $\delta$ was
introduced by Osaki (1996) precisely for obtaining good fits.

When heating of the outer parts of the disc is included
Eq. (\ref{eq:oi}) is no longer valid. Osaki (1996) assumed that in
quiescence the surface density is parallel to the critical density
profiles, that $\Sigma\sim 2 \Sigma_{\rm min}$, and he used the standard
critical densities. Things are different when heating of the outer disc
is included. Near the disc outer rim both $\Sigma_{\rm min}$ and
$\Sigma_{\rm max}$ are considerably reduced as is the distance between
them. These effects reduce the value of $\dot{M}_{\rm A-B}$.

This reduction is not sufficient to obtain desired values of the
mass-transfer rate, i.e. to $\sim 3 \times 10^{15}$ g s$^{-1}$, which is
typical of the mass transfer rate in SU UMa systems. The only hope is in
the strong dependence of $\dot{M}_{\rm A-B}$ on the outer disc radius.
The mean radius we have assumed for short period systems ($2.3 \times
10^{10}$ cm) is probably too large for most SU UMa's since it exceeds in
many cases the 3:1 resonance radius, which is of order of 0.46 $a$, i.e.
$2 \times 10^{10}$ cm for a total mass of 0.7 M$_\odot$ and a period of
1.6 hr characteristic of SU UMa stars. The 3:1 resonance radius should
be reached only during superoutbursts and should be much smaller (0.3
$a$, i.e. $1.3 \times 10^{10}$ cm) during the first few outbursts of a
supercycle . For such small radii $\dot{M}_{\rm A-B}$ would then be
reduced by a factor $\sim$ 4 -- 5, i.e. to a value slightly larger than
the estimated mass transfer rate in SU UMa systems.

\begin{figure}
\resizebox{\hsize}{!}{\includegraphics{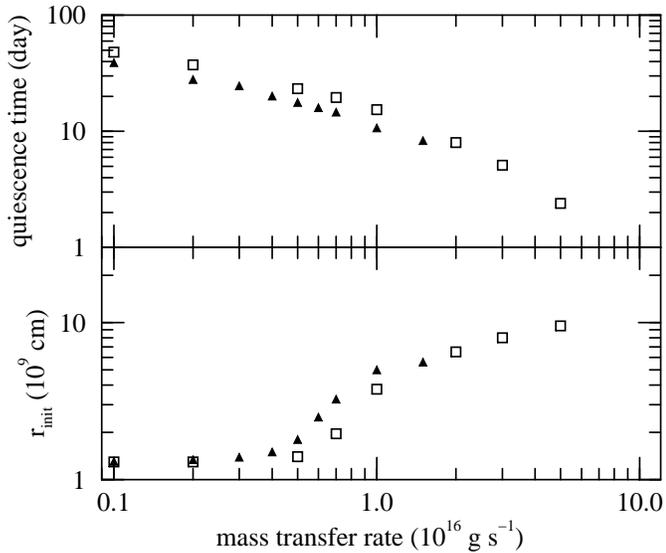}}
\caption{Radius $R_{\rm init}$ at which the instability is triggered and
quiescent time as a function of the mass transfer rate from the secondary.
The parameters are those of short period systems with $<r_{\rm out}> \sim 1.3
\times 10^{10}$ cm. Squares correspond to the standard DIM, and circles
to the case where the stream impact has been included.}
\label{fig9}
\end{figure}

In Fig. \ref{fig9} we show the quiescence time and the radius at which
the outburst starts versus mass transfer rate, for a short period system
with a mean outer disc radius $\sim 1.3 \times 10^{10}$ cm. This model
corresponds to a relaxed disc, i.e. a disc whose mass is constant over
an outburst cycle. Such small discs are expected to exist only during
the first half of a supercycle and later should grow in size. We have
assumed a large coefficient $c$ in Eq. \ref{eq:tid}, in order to obtain
a small disc, but we have not included heating by tidal torques that
would have resulted in much too large a dissipation: the dissipation of
tidal torques must be negligible in discs much smaller than the
truncation radius. As can be seen, outside-in outbursts occur for mass
transfer rates as low as $5 - 6 \times 10^{15}$ g s$^{-1}$, reasonably
close to the expected value in these systems.

It is worth noting that the recurrence time depends on the mass transfer
rate, even in the case of inside-out outbursts. A comparison of Figs.
\ref{fig4}, \ref{fig5} and \ref{fig9} shows that this effect is related
to the size of the disc. This is due to the fact that a local
dimensional analysis becomes invalid when the radius of the outer edge
of the disc is not much larger than the inner radius (for example, in
the case of steady discs, the inner boundary condition introduces a
factor $[1 -(r_{\rm in}/r)^{1/2}]$). Osaki (1996) asserts that the
correlation $t_{\rm s} \propto t_{\rm n}^{0.5}$ found between the
supercycle recurrence time $t_{\rm s}$ and the recurrence time of normal
outbursts $t_{\rm n}$ in SU UMa systems results from Eq.(\ref{tacc}) and
the relation $t_{\rm s} \propto \dot{M}^{-1}$ which results from the
assumption that super-outburst occur when sufficient mass is accumulated
(Osaki 1996; Hameury et al. 2000). Clearly, this cannot be the case if
one uses the standard disc model in which $\alpha$ is not forced to
adopt a functional dependence that would give the desired effect but is
just kept constant. In any case, the mass transfer rate cannot be
assumed to be the only variable on which recurrence times depend. For
example Menou (2000) showed that both the normal outburst and
super-outburst recurrence times strongly depend on the secondary to
primary mass ratio.

Small discs are also expected if one reduces the index $n$ in Eq.
(\ref{eq:tid}), since the smaller $n$, the larger the amplitude of disc
radius variations during outbursts; for a given maximum disc radius (some
fraction of the Roche radius), the disc radius during quiescence is therefore
expected to be smaller for smaller $n$.

\subsection{Observations}
\label{obs}

Campbell (1934) identified four distinct types of outburst in \object{SS
Cyg} light curves which have a large variety of characteristic rise
times. Since then, several techniques have been used to investigate the
physical outburst properties: multi-wavelength photometry, spectrometry,
eclipse mapping and model fitting. The first disc instability models
have predicted the outburst types A (outside in) and B (inside-out)
(\cite{smak84}), and showed that the first category rises earlier at
optical wavelengths and have a more asymmetric shape. As a result,
asymmetric outbursts are usually supposed to be of type A. However
asymmetric outbursts can also be of the inside-out type. For instance,
in Fig. \ref{fig6}, long outbursts are not symmetric but are of type B.

The UV delay is the observed lag between the rise in optical and the
rise in UV. Its duration is believed to be a signature of the outburst
starting region. In the naive picture since the hottest parts of the
disc associated with the shortest wavelengths are close to the white
dwarf, there should be almost no UV delay if the outburst starts in the
inner part of the accretion disc, whereas one should expect an important
UV delay for outside-in outbursts. However, as discussed by Smak
(1998), things are more complicated than what the naive picture would
suggest, and long delays can also exist for inside-out outbursts if the
outer disc radius expansion during outburst and the correct size of the
disc are taken into account. For a given binary system the UV delay will
always be longer for an outside-in outburst than for an inside-out one,
but the length of the delay is not sufficient by itself to determine the
outburst type.

Smak (1998) proposed that the ratio of the UV delay and the outburst duration
is a good indicator of the outburst type. However, the existence of both long
and short outbursts at a given orbital period makes this test very uncertain.
Indeed Fig. \ref{fig6} shows a probing counter-example where both short and
long outbursts are of type B and for which Smak's ratio would be quite
different.

In addition Hameury et al. (1999), using SS Cyg parameters, reproduced the
observed (\cite{mauche96}) asymmetric outburst and 1-day EUV delays for an
{\sl inside-out} outburst starting to propagate at the inner edge of a
truncated disc. This obviously raises the question of the real relationship
between the outburst type and the UV delay.

It has been known for some time that eclipse profiles allow to determine
the propagation direction of an outburst (Smak 1971 for U Gem; Vogt 1984
and Rutten, van Paradijs \&Tinbergen 1992 for OY Car). For example in 
the case of an accretion disc whose luminosity is much larger than both 
the white dwarf and the hot spot luminosity, the eclipse profile is almost 
symmetric. This is the case of \object{HT Cas}, a SU UMa system with a period 
of 106 minutes (\cite{ioanou99}). The authors found that during the rise of
the burst, the eclipse is shallow and its width is large. This means
that most of the flux originates from the outer parts of the disc. They
observed that the width of the eclipse decreases after the outburst
peak, which is in perfect agreement with the predicted decrease of the
disc radius during decline, therefore proving the outside-in nature of
the outburst. Webb et al. (1999) and Baptista, Catal\'an \& Costa (2000)
used the same technique to bring evidences of the inside-out nature of
an outburst of \object{IP Peg} and \object{EX Dra} respectively.

At present, the only way to be sure of the outburst type is to use, whenever
possible, eclipse profiles. As eclipses are not present in all systems it
would be worthwhile to test Smak's assumption on the UV delay in eclipsing
dwarf novae.

\section{Conclusions}

We have shown that the inclusion of additional heating effects such as
the stream impact and tidal dissipation might solve several problems of
the standard DIM.

The most important result is that when the tidal torque dissipation in the
outer disc regions is included, one obtains for a certain range of
mass-transfer rates an alternation of short and long outbursts which is
observed in long orbital period system such as \object{SS Cyg} (\cite{can92})
or \object{Z Cam} (\cite{oppen98}). The long outbursts are slightly brighter
than short outbursts as observed. Moderate fluctuations of the mass-transfer
rate would extend the range for which such alternating outbursts are present.
For systems with mass transfer rates close to the stability limit, such
fluctuations will produce standstills as shown in Buat-M\'enard et al.
(2000). This solution of the long/short outburst problem (Smak 2000) is
different from the one suggested by Smak (1999a). In our model mass-transfer
fluctuations play only an auxiliary role and long outbursts do not result
from fluctuations bringing the disc to a steady state.

We showed that additional heating effects make the outside-in outburst
possible at low mass transfer rates without changing the
$\alpha$-prescription. Our results on mass transfer rates and outburst types
are in good agreement with most estimates in the case of long period
systems. In the case of short period systems of the SU UMa type, for
outside-in outburst to occur the disc would have to be smaller than usually
assumed. For small discs the recurrence times of both the inside-out and
outside-in outbursts is $\dot{M}_2$-dependent. It must be stressed out that
there are still many uncertainties on the observational determination of
outburst types. For instance, the UV delay is not necessarily a good
criterion for determining the outburst type; the eclipse profile during
outburst is a much better criterion, but unfortunately restricted to a small
subset of dwarf novae.

There are still several points that need to be clarified. First, the stream
impact effect geometry is not well known and is considered here in a 1D
approximation (as for the tidal dissipation). We do not know how far and how
efficiently the energy propagates in the disc. The expression of the tidal
torque can also be subject to discussion. Moreover, we did not include
irradiation here, and it would be interesting to study the combined effects
of all phenomena. Finally, as raised above, the existing discrimination
criteria on the outburst types A and B have to be tested and improved.

\begin{acknowledgements}
We thank Guillaume Dubus for helpful comments.
This work was supported in part by a grant from {\sl Programme National de
Physique Stellaire} of the CNRS.
\end{acknowledgements}

\listofobjects

\begin{thebibliography}{}

\bibitem[]{}Armitage P.J., Livio M., 1996, ApJ 470, 1024

\bibitem[Armitage \& Livio 1998]{arm}Armitage P.J., Livio M., 1998, ApJ
493, 898

\bibitem[Baptista et al. 2000]{bcc} Baptista R.,
        Catal\'an M.S., Costa L., 2000, MNRAS, 316, 529 

\bibitem[Baraffe \& Kolb 1999]{bk1999}Baraffe I., Kolb U., 1999, MNRAS 309, 1034

\bibitem[Buat-M\'enard et al. 2000]{buat2000}Buat-M\'enard V., Hameury J.-M.,
Lasota J.-P., 2000, A\&A submitted

\bibitem[Campbell 1934]{cam34}Campbell L., 1934, AHCO 90, 93

\bibitem[Cannizzo 1993]{can93}Cannizzo J.K. 1993, ApJ 419, 318

\bibitem[Cannizzo \& Mattei 1992]{can92}Cannizzo J.K., Mattei J.A. 1992, ApJ
401, 642

\bibitem[Cannizzo, Wheeler \& Polidan 1986]{can86}Cannizzo J.K., Wheeler
J.C., Polidan R.S., 1986, ApJ 301, 634

\bibitem[Dubus et al. 1999]{dlhc99}Dubus G., Lasota J.-P., Hameury J.-M.,
Charles P., 1999, MNRAS 303, 139

\bibitem[Hameury, Lasota \& Hur\'e 1997]{jmh97b}Hameury J.-M., Lasota J.-P.,
Hur\'e J.-M., 1997, MNRAS 287, 937

\bibitem[Hameury et al 1998]{jmh98}Hameury J.-M., Menou K., Dubus G., Lasota
 J.-P., Hur\'e J.-M., 1998, MNRAS 298, 1048

\bibitem[Hameury, Lasota \& Dubus 1999]{jmh99}Hameury J.-M., Lasota J.-P.,
Dubus G., 1999, MNRAS 303, 39

\bibitem[Hameury, Lasota \& Warner 2000]{jmh00}Hameury J.-M., Lasota J.-P.,
Warner B., 2000, A\&A 353, 244

\bibitem[Harrop-Allin \& Warner 1996]{hw96}Harrop-Allin M.K., Warner B.,
1996, MNRAS 279, 228

\bibitem[Ichikawa \& Osaki 1992]{ich-osak92}Ichikawa S., Osaki Y., 1992, PASJ
44, 15

\bibitem[Ichikawa \& Osaki 1994]{ich-osak94}Ichikawa S., Osaki Y., 1994, PASJ
46, 621

\bibitem[Ioannou et al. (1999)]{ioanou99}Ioannou Z., Naylor T., Welsh W.F.,
Catal\'an M.S., Worraker W.J., James N.D., 1999, MNRAS 310, 398

\bibitem[Lasota 2000]{lasota2000} Lasota J.-P., 2000, New Astronomy Reviews,
submitted

\bibitem[Lasota, Hameury \& Hur\'e 1995]{lasota95} Lasota J.-P., Hameury
J.-M., Hur\'e J.-M., 1995, A\&A 302, L29

\bibitem[Livio \& Pringle 1992]{livio92}Livio M., Pringle J., 1992, MNRAS 259,
23

\bibitem[Lubow \& Shu 1975]{lub-shu75}Lubow S.H., Shu F.H., 1975, ApJ 198, 383

\bibitem[Marsh \& Horne 1990]{marsh90}Marsh T.R., Horne K., 1990, ApJ 349, 593

\bibitem[Mauche 1996]{mauche96}Mauche C.W., 1996, in Astrophysics in the Extreme
Ultraviolet, eds.\ S. Bowyer and R.F. Malina, Dordrecht, Kluwer, p.\ 317.

\bibitem[Menou 2000]{m2000}Menou K., 2000, Science, in press

\bibitem[Meyer \& Meyer-Hofmeister 1983]{meyer83}Meyer F., Meyer-Hofmeister E.
1983, A\&A 121, 29

\bibitem[Meyer \& Meyer-Hofmeister 1994]{meyer94}Meyer F., Meyer-Hofmeister E.
1994, A\&A 228, 175

\bibitem[Oppenheimer, Kenyon \& Mattei 1998]{oppen98}Oppenheimer B.D., Kenyon
S.J., Mattei J.A., 1998, AJ 115, 1175

\bibitem[Osaki 1985]{o85}Osaki Y., 1985, A\&A 144, 369

\bibitem[Osaki 1989]{osak89}Osaki Y., 1989, PASJ 41, 1005

\bibitem[Osaki 1996]{o96}Osaki Y., 1996, PASP 108, 39

\bibitem[Paczy\'nski 1977]{pacz77}Paczy\'nski B., 1977, ApJ 216, 822

\bibitem[Papaloizou \& Pringle 1977]{pap-pring77}Papaloizou J., Pringle J.E.,
1977, MNRAS, 181, 441

\bibitem[Patterson 1984]{pat84}Patterson J., 1984, ApJS 54, 443


\bibitem[Ritter \& Kolb 1998]{ritter98}Ritter H., Kolb U., 1998, A\&AS 129, 83

\bibitem[Rutten et al. 1992]{rutt1992} 
Rutten R.G.M., van Paradijs J., Tinbergen J., 1992, A\&A, 260, 
213 

\bibitem[Shakura \& Sunyaev 1973]{shak73}Shakura N.I., Sunyaev R.A. 1973, A\&A
24, 337

\bibitem[Smak 1971]{smak71}Smak J., 1971, Acta Astr. 21, 15

\bibitem[Smak 1984]{smak84}Smak J., 1984, Acta Astron. 34, 161

\bibitem[Smak 1998]{smak98}Smak J., 1998, Acta Astron. 48, 677

\bibitem[Smak 1999a]{smak99a}Smak J., 1999a, Acta Astron. 49, 383

\bibitem[Smak 1999b]{smak99b}Smak J., 1999b, Acta Astron. 49, 391

\bibitem[Smak 2000]{smak2000}Smak J., 2000, New Astronomy Reviews 44, 171

\bibitem[Spruit \& Rutten 1998]{spruit98}Spruit H.C., Rutten R.G.M. 1998,
MNRAS, 299 768

\bibitem[Stehle \& King 1999]{sk99}Stehle R., King A.R., 1999, MNRAS 304, 698


\bibitem[van Paradijs 1983]{vanP83}van Paradijs J., 1983, A\&A 125, L16

\bibitem[Vogt 1983]{vogt83}Vogt N., 1983, A\&A, 128, 29

\bibitem[Warner 1994]{w94}Warner B., 1994, Ap\&SS 226, 187

\bibitem[Warner 1995]{w95}Warner B., 1995, Cataclysmic Variable Stars.
Cambridge University Press, Cambridge

\bibitem[Webb et al. 1999]{webb99} Webb N.A., Naylor T., Ioannou Z.,
Worraker W.J., Stull J., Allan A., Fried R., James N.D., Strange D.,
1999, MNRAS 310, 407

\end{thebibliography}
\end{document}